\documentclass[sigconf, screen]{acmart}

\PassOptionsToPackage{dvipsnames}{xcolor}

\PassOptionsToPackage{colorlinks=true,linkcolor=RubineRed,breaklinks=True,citecolor=RubineRed}{xcolor}
\usepackage{enumitem}
\setlist[itemize]{noitemsep, topsep=0pt}

\usepackage[most]{tcolorbox}

\usepackage{adjustbox}
\usepackage[linesnumbered,algo2e,ruled]{algorithm2e}
\usepackage{multirow}
\usepackage{booktabs,subcaption,dcolumn}
\usepackage{tikz}
\usepackage{enumitem}
\usepackage{tabularx}

\usepackage{pifont}
\usepackage{svg}
\usepackage{soul}

\usepackage{amsmath}
\usepackage{amsfonts} 

\usepackage{colortbl}

\newcolumntype{?}{!{\vrule width 1.5pt}}

\newtcolorbox{cooltextbox}[1][]{%
    colback=black!5,
    colframe=black!5,
    notitle,
    sharp corners,
    borderline west={1pt}{0pt}{red!80!black},
    enhanced,
    breakable,
    }
\newcommand{\element}[1]{\textsc{#1}}

\newcommand\smabb[1]{{\small $\mathbb{#1}$}}
\newcommand\scbb[1]{{\scriptsize $\mathbb{#1}$}}

\newcommand\revision[1]{%
  \bgroup
  \hskip0pt\color{black}%
  #1%
  \egroup
}

\DeclareUnicodeCharacter{0306}{ }

\newtheorem{definition}{\textsc{Def.}}

\newtheorem{remark}{\textsc{Remark}}

\AtBeginDocument{%
  \providecommand\BibTeX{{%
    \normalfont B\kern-0.5em{\scshape i\kern-0.25em b}\kern-0.8em\TeX}}}

\setcopyright{none}

\acmConference[ICSS '22]{Annual Industrial Control System Security Workshop}{December 6, 2022}{Austin, TX, USA}
\acmBooktitle{Proceedings of the 8th Annual Industrial Control System Security Workshop (ICSS '22), December 6, 2022, Austin, TX, USA}

\begin{document}
\title{Cybersecurity in the Smart Grid: Practitioners' Perspective}

\author{Jacqueline Meyer}
\orcid{0000-0001-7135-2052}
\email{jacqueline.meyer@uni.li}
\affiliation{%
  \department{Institute of Information Systems}
  \institution{University of Liechtenstein}
  \country{}
}

 \author{Giovanni Apruzzese}
 \orcid{0000-0002-6890-9611}
\email{giovanni.apruzzese@uni.li}
\affiliation{%
    \department{Institute of Information Systems}
  \institution{University of Liechtenstein}
  \country{}
}

\begin{abstract}
The Smart Grid (SG) is a cornerstone of modern society, providing the energy required to sustain billions of lives and thousands of industries. Unfortunately, as one of the most critical infrastructures of our World, the SG is an attractive target for attackers. The problem is aggravated by the increasing adoption of digitalisation, which further increases the SG's exposure to cyberthreats. Successful exploitation of such exposure leads to entire countries being paralysed, which is an unacceptable---but ultimately inescapable---risk. 

This paper aims to mitigate this risk by elucidating the perspective of real practitioners on the cybersecurity of the SG. We interviewed 18 entities, operating in diverse countries in Europe and covering all domains of the SG---from energy generation, to its delivery. 
Our analysis highlights a stark contrast between (a)~research and practice, but also between (b)~public and private entities. For instance: some threats appear to be much less dangerous than what is claimed in related papers; some technological paradigms have dubious utility for practitioners, but are actively promoted by literature; finally, practitioners may either under- or over-estimate their own cybersecurity capabilities.
We derive four takeaways that enable future endeavours to improve the overall cybersecurity in the SG. We conjecture that most of the problems are due to an improper communication between researchers, practitioners and regulatory bodies---which, despite sharing a common goal, tend to neglect the viewpoint of the other `spheres'.
\end{abstract}


\settopmatter{printfolios=true}

\maketitle


\section{Introduction}
\label{sec:introduction}
Among the infrastructures that sustain the modern world, one in particular stands out: the Smart Grid (SG). Tasked to provide the energy empowering our society, without the SG most services, commodities, and advances, would be either significantly impaired, or simply impossible to deliver~\cite{dileep2020survey}.

\begin{figure}[!htbp]
    \centering
    \includegraphics[width=0.99\columnwidth]{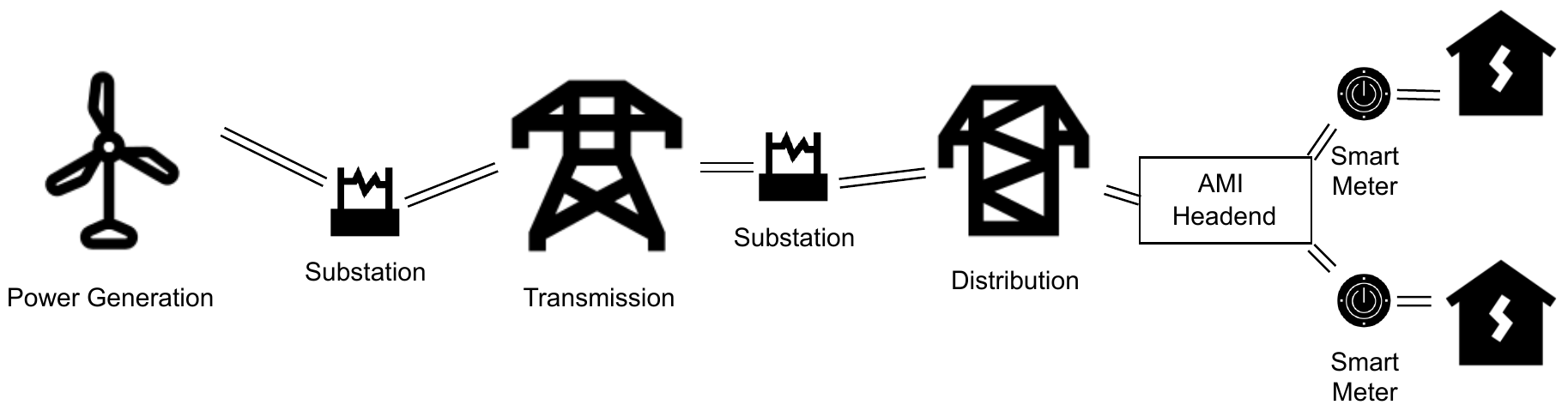}
    \caption{Overview of the Smart Grid.}
    \label{fig:overview}
\end{figure}

We provide a schematic representation of some exemplary elements\footnote{All such elements can be considered as Cyber-Physical Systems (CPS)~\cite{PC_krotofil2014cps} or part of Industrial Control Systems (ICS)~\cite{PC_ayub2021empirical}.} of the SG in Fig.~\ref{fig:overview} (which is our adaptation from~\cite{el2018smartgrid}). After some energy is generated at a given source, the SG must transmit such energy (in the form of electricity) to various devices which ensure the proper energy distribution to the end-users. Given its strategical importance, the SG is increasingly relying on Information Technology (IT) to further enhance its functionalities~\cite{peng2019survey}. For example, IT improves the reliability~\cite{aydeger2019sdn_CPSW} and efficiency of the SG~\cite{singh2021end}, and facilitates the collection and distribution of energy in remote areas~\cite{mendicino2019corporate}, or in resource-constrained settings~\cite{hanna2021efficient_CPSW}.

Unfortunately, the SG is well-known to be a preferred target for attackers~\cite{PC_rowe2015attribution}, and reliance on IT inevitably exposes to the risk\footnote{A recent report~\cite{swissPGrisk} quantifies such risk, stating that attacks against the Swiss SG can cause losses of up to 12 billion CHF ($2\%$ of Swiss' GDP).} of cyberthreats~\cite{al2016security}.
Early cyberattacks date back to 2003, when the David-Besse nuclear power plant in the USA was affected by the well-known Slammer malware~\cite{alladi2020industrial}. Other notable examples include famous Advanced Persistent Threats (APT), such as Stuxnet in 2006~\cite{Virvilis:BigFour} or the attack to the Ukrainian SG in 2015~\cite{case2016analysis}. The latter, in particular, caused outages to over 200K households as a result of the compromise of three major country-wide energy suppliers. To safeguard the correct operation of the SG, it is paramount to constantly improve its cybersecurity---which is a topic covered by abundant literature (e.g.,~\cite{peng2019survey, el2018smartgrid, PC_mashima2021securing}).

Our paper is inspired by two recent works by Kumar et al.~\cite{kumar2020adversarial} and Grosse et al.~\cite{grosse2022so}. Despite focusing on a different context, these works highlight a stark ``disconnection'' between (a)~the claims made by researches and (b)~the viewpoint of real practitioners. Indeed, scientific papers tend to make assumptions that deviate from real-world scenarios---typically, due to the lack of information on how real IT systems work. Such a lack is even more common in critical infrastructures~\cite{PC_awad2018tools}, because any information leak can be exploited by attackers for their offensive campaigns~\cite{taylor2017security, PC_rowe2015attribution}. Simply put, many papers focus on issues that, from the practitioners' perspective, have unclear relevance to real systems. As we will show, this gap is present also in the SG context---which is further complicated by the regulations that govern the complex relationships of the SG ecosystem.
It is well-known that resources are limited in cybersecurity~\cite{apruzzese2022role}, and hence priority should be given to the most relevant and impactful issues---but only if such issues are brought to light. We aim to rectify this problem.

\textsc{\textbf{Our Contribution.}} 
This paper bridges the gap between research and practice in the SG context, with the intention of improving the cybersecurity of real SG systems.
To reach our objective, we begin by summarizing the limitations of existing literature from a `practical' viewpoint (§\ref{sec:related}). Then, we make three major contributions.
\begin{itemize}
    \item We \textit{conduct an extensive survey with real practitioners involved in the SG's cybersecurity} (§\ref{sec:research}). Our survey elucidates the viewpoint of 18 entities, spanning across all seven domains of the SG, and operating in diverse countries in Europe. Our questions cover generic cybersecurity aspects, e.g.: risk assessment, dangerous threats, utility of recent technologies.
    
    \item After transparently presenting our major findings (§\ref{sec:findings}), we perform an objective analysis \textit{highlighting the disconnections that emerge from our survey} (§\ref{sec:analysis}). We show: the discrepancy between research and practice (§\ref{ssec:pravres}); and the differences between the public and private sector (§\ref{ssec:pubvpriv}). 
    
    \item We then provide an \textit{original interpretation} of our results~(§\ref{sec:interpretation}). We explain the role of \textit{regulations} in operational cybersecurity (§\ref{ssec:regulations}); and derive \textit{takeaways} for the four `spheres' contributing to the cybersecurity of the SG (§\ref{ssec:takeaways}): companies, legislative bodies, researchers, and authorities.
\end{itemize}
To the best of our knowledge, this is the first paper that provides such an holistic coverage of `practical' cybersecurity in the European SG in the recent years.
\section{Motivation and Related Work}
\label{sec:related}
Many papers investigated various cybersecurity aspects of the SG. We identify four categories of related works: \textit{novel attacks and defenses}, \textit{literature reviews}, \textit{case studies}, and \textit{interviews}. Let us explain the necessity of our study by comparing our paper with prior work.

\textbf{Attacks and Defenses.}
Proposing novel attack scenarios, as well as corresponding countermeasures, is common in research. Yet, all such evaluations are performed through simulations, and therefore have \textit{poor practical value}. For instance, Zuo et al.~\cite{zuo2020unbounded} propose unbounded attacks on microgrids: despite being rooted on sophisticated mathematical foundations, the assessment is carried out in a hardware-in-a-loop testbed. Rrushi et al.~\cite{PC_rrushi2022physics} propose a physics-driven approach to counter CPS malware: although the reference data is collected from real substations, the experiments are carried out in a synthetic environment. A similar issue also affects other attacks, such as False Data Injection (FDI)~\cite{deng2018false, PC_yoo2019overshadow, liang2016review}, Denial of Service (DoS)~\cite{PC_krotofil2014cps, lu2020adaptive}, spoofing~\cite{chauhan2021synchrophasor}, or Man-in-the-Middle~(MitM)~\cite{PC_wlazlo2021man}. Put simply: no system is foolproof, and it is positive that research papers also investigate similar scenarios. However, practitioners have limited resources: according to the cyber-resilience best practices~\cite{conklin2017cyber,noel2017big}, such resources should be spent on threats that are more likely to endanger the real SG (which cannot be gauged through `attack/defense' papers). 

\textbf{Reviews.} Most reviews are \textit{exclusively based on scientific papers}. For instance, Awad et al.~\cite{PC_awad2018tools} focus on techniques for digital forensics in SCADA systems, and although they cover frameworks, methodologies, and implementations, all such considerations are based on past scientific literature. Furthermore, Peng et al.~\cite{peng2019survey} provide an in-depth analysis and identify some cybersecurity challenges in the SG, but their main focus is on specific threats (e.g., DoS and FDI), preventing a holistic coverage. Such limited scope is addressed, e.g., by El et al.~\cite{el2018smartgrid}, which also provide actionable recommendations. However, all findings of~\cite{peng2019survey, el2018smartgrid} are purely theoretical or derived from prior research, overlooking the insight of practitioners. 

\textbf{Case Studies.} Many papers provide exhaustive analyses on real APT targeting the SG. For instance, the authors of~\cite{al2018stuxnet} considers the evolutions of the original Stuxnet, while Case et al.~\cite{case2016analysis} focus on the Ukrainian SG. A more recent overview of reported APT is provided by Kaura et al.~\cite{kaura2022analysing}. These papers are useful to provide some practical takeaways; however, \textit{they focus on attacks launched many years prior}, and do not allow to assess the current state-of-the-art of cybersecurity in the SG.

\textbf{Interviews.}
\textit{Few works directly interviewed practitioners}, and especially those related to the SG. For instance, Fischer et al.~\cite{fischerHubner2021stakeholder} focus on general cybersecurity challenges in critical infrastructures. Despite providing insights derived from 63 stakeholders, such experts pertain to different contexts (e.g., Smart Cities) than the SG. Some papers report `outdated' findings---e.g., Line et al.~\cite{line2013case} carry out 19 interviews, but in in 2012. Nonetheless, Siemens et al.~\cite{siemers2021modern} conduct a ``workshop'' in 2021 entailing participants (23 in total) from industry and academia, there is no information about the composition of these two groups, preventing to distill meaningful knowledge of the practitioners' viewpoints. 
Perhaps the closest effort to our paper is~\cite{randall2021cybersecurity}, which focuses on the perspective of 10 organizations in the power industry. However, their findings only span across a single country (the US), and only focus on information sharing---preventing a broad coverage of the topic.

\begin{cooltextbox}
\element{\textbf{Research Gap.}} 
Existing works do not provide a holistic vision of the current cybersecurity in the SG from the perspective of practitioners. Our paper aims to fix this gap by elucidating the recent opinion of SG experts (operating in diverse countries in Europe) to the research domain.
\end{cooltextbox}

\section{Research Methodology}
\label{sec:research}
The main contribution of this paper are the findings of our survey with 18 entities related to the SG. Our findings, and corresponding survey, revolve around a broad research objective: \textit{investigating the state-of-the-practice of cybersecurity in the European SG}.

\begin{figure}[!htbp]
    \centering
    \includegraphics[width=0.99\columnwidth]{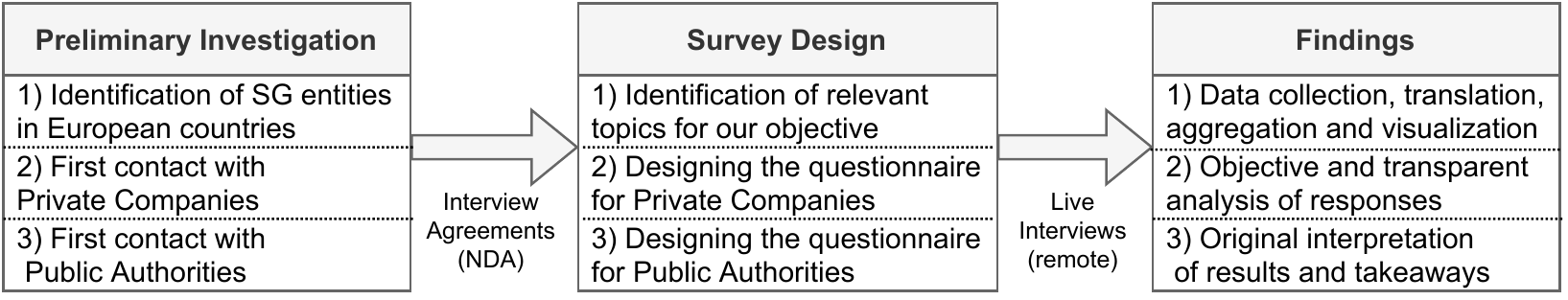}
    \caption{Overview of our adopted research methodology.}
    \label{fig:method}
    \vspace{-0.5cm}
\end{figure}

In what follows, we describe our research methodology---for which we provide an overview in Fig.~\ref{fig:method}. At the end of this section, we make some considerations on our study and elucidate some of the challenges we encountered (§\ref{ssec:challenges}).

\subsection{Preliminary Investigation}
\label{ssec:preliminary}

We began our study by identifying a suitable set of entities that (i)~allowed to address our main objective and that (ii)~were willing to contribute to our research. Let us briefly summarize the complexity of the modern SG, so as to enable understanding why reaching our objective is difficult for research endeavours.

\begin{figure}[!htbp]
    \centering
    \includegraphics[width=0.99\columnwidth]{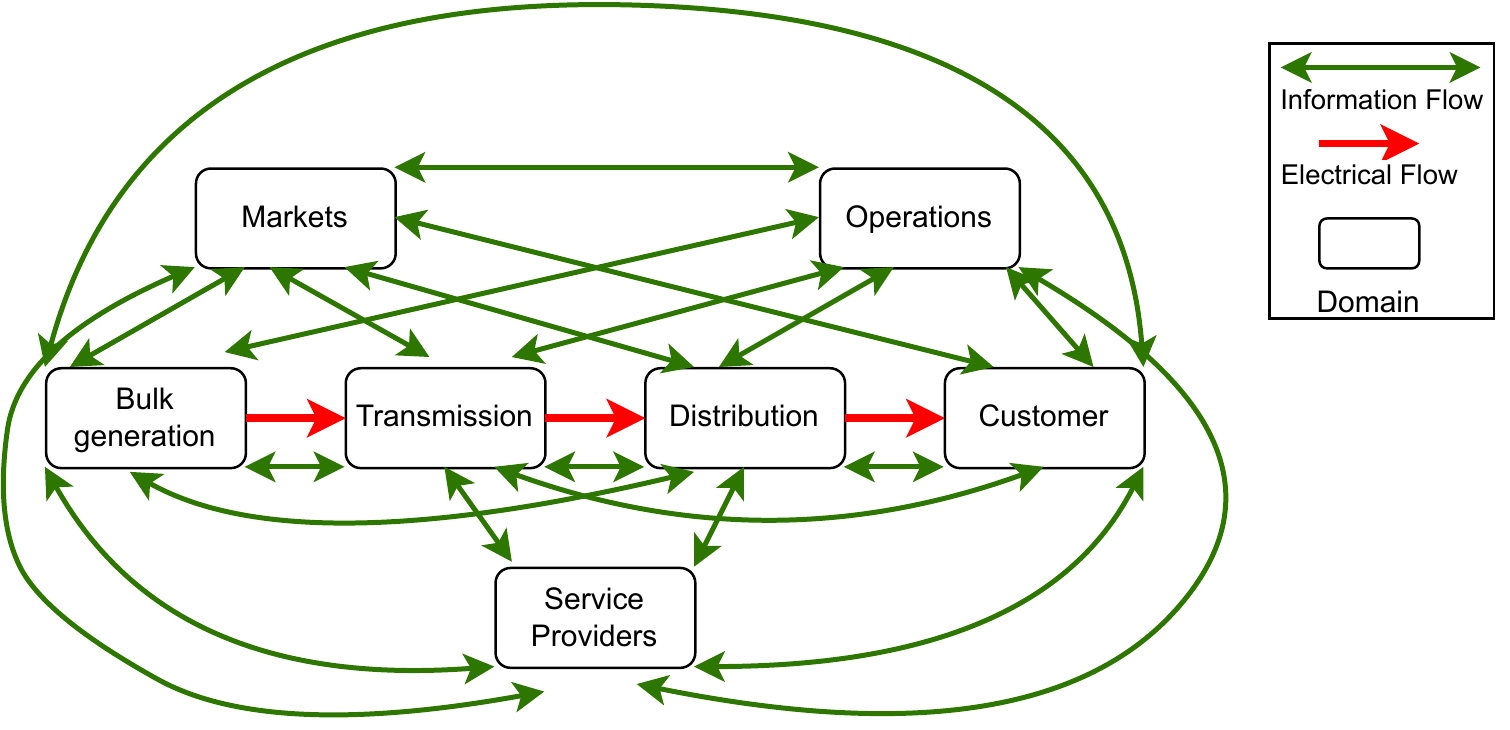}
    \caption{The NIST conceptual model of the SG, spanning across 7 domains---all of which are covered in our research.}
    \label{fig:concept}
\end{figure}

\textbf{Background.} The SG is a network of components (see Fig.~\ref{fig:overview}) that, despite having the same underlying goal (i.e., delivering energy from a source to a destination), are owned, managed and operated by diverse entities. In particular, a common way to view the SG is through the NIST conceptual model~\cite{fitzpatrick2010nist}, schematically depicted in Fig.~\ref{fig:concept}. It identifies seven interconnected domains which have a crucial role in the SG network. All such domains relate to each other---either directly (via electrical flows) or indirectly (via, e.g., functional dependencies or market demands). Simply put, the SG is a complex ecosystem in which each domain solves a specific function---hence, even a single failure can impair the entire system. For this reason, we conduct our research by interviewing entities \textit{from all seven SG domains}. 
Such design allows our study to overcome the limitations of past work and provide a holistic understanding of the current state of the SG from a cybersecurity viewpoint. 

\textbf{Interviewed Entities.}
After contacting dozens of entities related to the SG in diverse countries in Europe, we reached an `interview agreement' with 18 entities, which we divide in two groups.
\begin{itemize}
    \item \textit{Private Companies.} We reached an agreement with 14 private companies covering the `operational' domains of the SG, namely: Bulk Generation, Transmission, Distribution, Operations, Markets, Service Providers. Moreover, these companies are intertwined via business relationships, hence they can also be considered as part of the Customer domain. All companies operate the SG in diverse countries in Europe. 
    \item \textit{Public Authorities.} We reached an agreement with 4 public authorities, representing the `final-customer' domain of the SG. The idea is providing a complementary perspective whose focus is not on the ``actual security of the SG'', but rather on the ``perceived security of the SG''. Such public authorities respond to the European countries in which the 14 private companies have their main headquarters.
\end{itemize}
For simplicity, in the remainder we will use \smabb{C} to denote (private) \textit{companies}, and \smabb{A} to denote (public) \textit{authorities}. Due to NDA, we cannot reveal further information on either \smabb{A} or \smabb{C}.

\subsection{Survey and Interviews}
\label{ssec:interviews}
The next step after finding an agreement was designing a set of questions (i.e., a \textit{questionnaire}) that (i) allowed to reach our objective, but which (ii) could be answerable by our interviewees. 

\textbf{Scope and Questions.}
Given the heterogeneity of our population (which cover all SG domains), we only consider \textit{high-level} questions pertaining to cybersecurity. Moreover, we follow established practices (e.g.,~\cite{fischerHubner2021stakeholder}) and thereby our interviews are mostly \textit{structured}. Such format is appropriate because it allows to derive \textit{quantitative} results, and it protects against possible violations of NDA. 
In particular, we asked 10 closed questions to \smabb{A}; and 30 questions to \smabb{C}, split into 27 closed questions with pre-defined answers, and 3 open questions---hence resembling a semi-structured~\cite{leech2002asking} interview design (as also done in~\cite{line2013case}). The 27 closed questions are the same for all \smabb{C}; whereas the 3 open questions are tailored for the specific domain of a given \smabb{C}. 
Overall, our questions span across several generic cybersecurity aspects, e.g.: experience with past attacks, risk-assessment, utility of novel technologies, and the most challenging cyberthreats\footnote{The questionnaires containing all closed questions to \scbb{C} and \scbb{A} are in this repository: \url{https://github.com/hihey54/smartgrid_survey}.}. 

\textbf{Interviews.}
The interviews consisted in remote one-on-one meetings, conducted between one of the authors and a representative person (with technical expertise) of the corresponding entity. During such meetings, we asked the questions and showed the possible answers to the respondent, granting them the option of not answering\footnote{The interviews were done with single individuals, who -- despite their technical expertise -- may not have had the necessary knowledge to provide an exhaustive answer (or simply could not respond due to NDA).}. 
The interviews with \smabb{C} lasted {\small$\sim$}1~hour, whereas those with \smabb{A} lasted {\small$\sim$}30 minutes. 
During the interviews we took plenty of notes---amounting to more than 200k characters overall. We used such notes to derive our own interpretations of our findings (§\ref{sec:interpretation}).

\textbf{Timeline.} An important remark concerns the timeline of our survey. We interviewed \smabb{C} from Jan.~2022 until Feb.~2022, i.e., \textit{before} the breakout of the Russian/Ukrainian conflict~\cite{serpanos2022cyberwarfare}. On the other hand, we interviewed \smabb{A} in Mar.~2022, i.e., \textit{after} the conflict erupted. Such an event was not planned, and although there is a slight chance that it may have affected the responses of \smabb{A}, we emphasize that we treat the two groups as independent from each other. Therefore, our results are consistent across all of \smabb{C} and all of \smabb{A}.

\subsection{Correctness and Challenges}
\label{ssec:challenges}
Before presenting our results, let us make some insightful remarks, with the twofold purpose of (i)~validating our efforts; and (ii)~providing a source of inspiration for future studies.

\textbf{Preventing Bias.}
All our interviews and subsequent analyses have been carried out so as to minimize the introduction of bias. 
\begin{itemize}
    \item The \textit{questionnaire} was designed by two authors, who had frequent meetings aimed at formalizing questions that could be answered by each of the considered entities (i.e., \smabb{C} or \smabb{A}).
    \item During the \textit{interviews} (done by one of the authors), participants were explicitly asked to respond objectively, and to clearly state whether they were unsure of their answers. In case of uncertainty, we still registered their response (if given), but we then used our notes to derive the correct option by weighing the individual response against those given by other similar entities. If -- even after analysing our notes -- we could not infer a clear response, we assigned the response to the `not provided' category to avoid generating noise.
    \item During the \textit{analysis} (done by one of the authors), regular breaks were taken to fairly interpret the answers received during the interviews~\cite{blair2015reflexive}. Such analyses have been repeated three times to account for the possibility of honest mistakes. At the end of this process, two of the authors discussed the findings to reach a consensus.
\end{itemize}
We can safely assume that our results are scientifically correct.

\textbf{Challenges.}
We find instructive to report some details about the difficulties we encountered: getting in touch with companies is a challenging endeavour for researchers.
\begin{itemize}
    \item We initially aimed to interview more than 30 private companies, but ultimately only 14 accepted: some were unresponsive, or were not willing to release any kind of information (even under NDA). 
    
    \item In contrast, the 4 public authorities (which we contacted \textit{after} finding some agreements with the private companies) were more willing to cooperate.
    
    \item Out of the 14 private companies, 5 agreed to help us only after phone calls lasting more than 60 minutes.
    
    \item Only 5 of the interviews with the 14 private companies were carried out on the initial scheduled date (many had to be postponed at the last minute).
    
    \item To set up the interviews with all our entities, we sent a total of 145 emails between Nov. 2021 and Feb. 2022. 
    
    \item Overall, our entities pertain to diverse countries---hence, the interviews were carried out in the interviewee's official language. This required us to (i) create language-specific versions of our questionnaire; and (ii) translate all responses into English before analyzing them.
\end{itemize}
All such difficulties make us believe that the `gap' between research and practice could be reduced if (private) companies were more willing to cooperate with academia. We acknowledge, however, that companies dealing with critical infrastructures (such as the SG) have valid reasons for not disclosing sensitive security information.

\section{Major Findings}
\label{sec:findings}
We now present the most insightful findings\footnote{Due to space limitations, reporting and analyzing \textit{all} questions of our survey is not possible within this single paper. Upon request, however, we are willing to disclose a `question-by-question' analysis of our survey---alongside all the operations we followed to derive our results. Our repository also contains the source-code of all our graphs.} of our survey. In particular, we report those results that: (i)~allow to gauge the state-of-practice of cybersecurity in the SG---described in this section; but that also (ii)~reveal the presence of some \textit{blind spots}. The latter---which we objectively analyse (§\ref{sec:analysis}) and attempt to interpret (§\ref{sec:interpretation})---represent a valuable avenue for future work. (Due to NDA, we provide the aggregated results for \smabb{C} and \smabb{A}: fine-grained analyses could be used to identify some of our respondents.)


\subsection{Overview} 
\label{ssec:main}
\begin{figure}[!htbp]
    \centering
    \includegraphics[width=0.8\columnwidth]{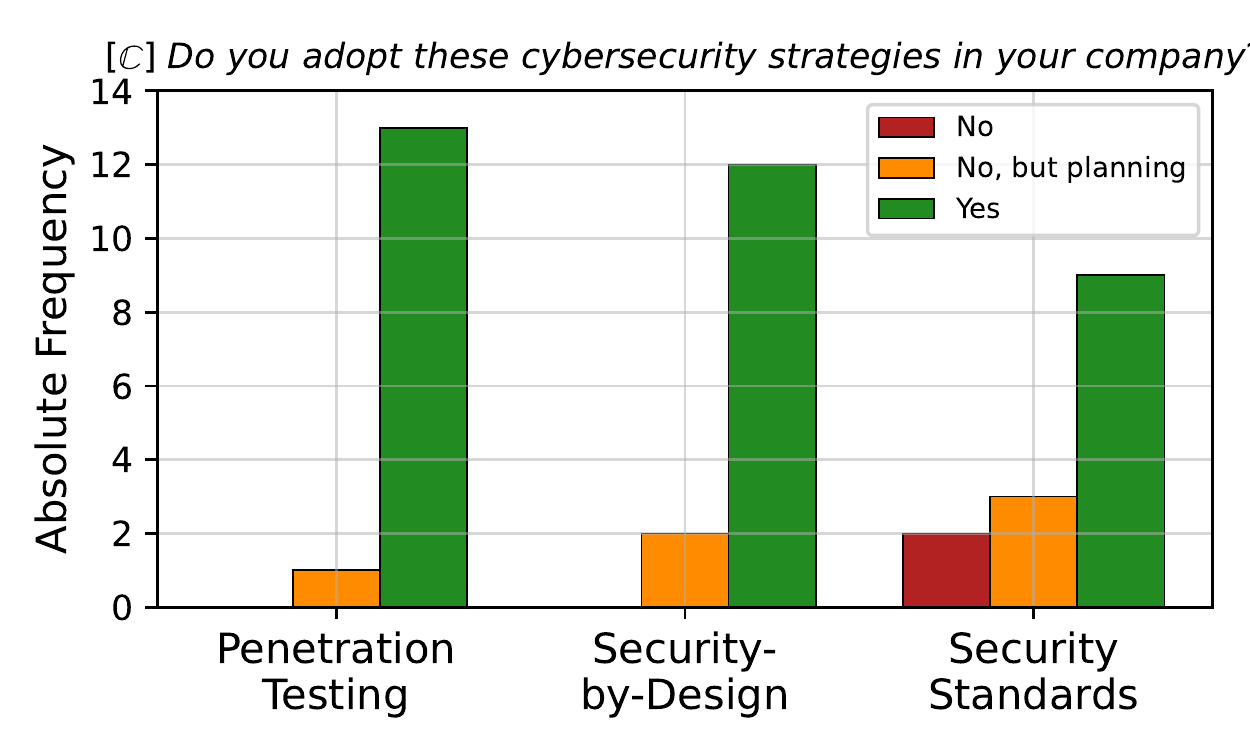}
    \caption{Answers of \scbb{C} on their adoption of security \textit{strategies}.}
    \label{fig:strategy}
\end{figure}

Some of our findings confirm what is already well-known. For instance, 64\% of \smabb{C} (reportedly) were \textit{targeted by cyberattacks} in the last decade. 
Only 20\% of \smabb{C} entirely manage their cybersecurity, while 80\% \textit{outsources} it at varying degrees. All \smabb{C} adopt both \textit{backup} and \textit{data-replication} strategies.
Moreover, \smabb{C} considers that \textit{data confidentiality} to be problematic, with an average of 8 on a [1--10] scale (higher is more problematic).
We report in Fig.~\ref{fig:strategy} the viewpoint of \smabb{C} with respect to established cybersecurity strategies; perhaps surprisingly, two companies do not (nor plan to) integrate any security standard (e.g., the ISO27001~\cite{fitzpatrick2010nist}).

\vspace{-0.5em}

\begin{figure}[!htbp]
    \centering
    \includegraphics[width=0.8\columnwidth]{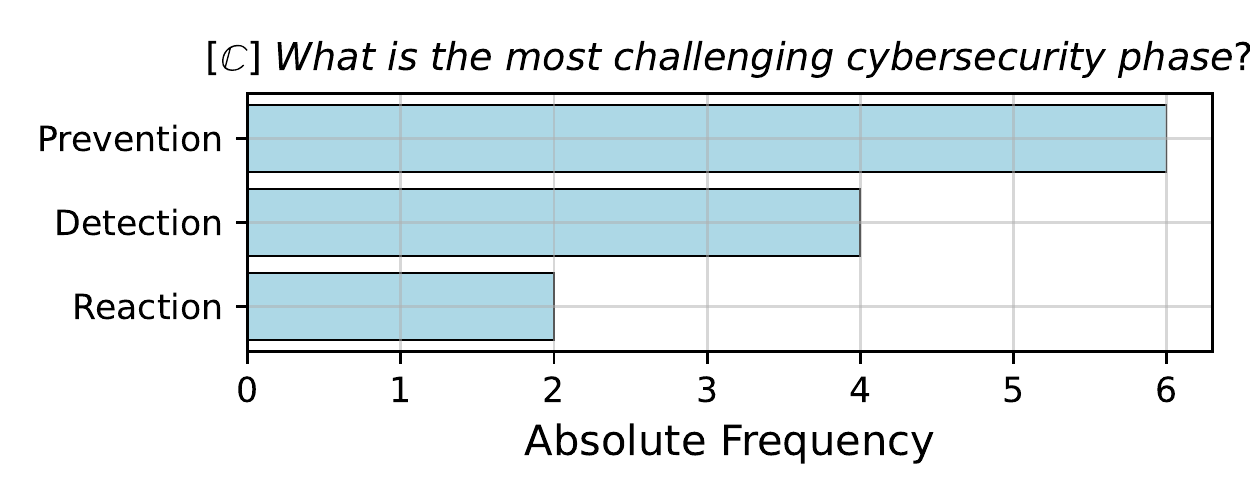}
    \caption{Most challenging phases of cybersecurity (for \scbb{C}).}
    \label{fig:phases}
\end{figure}

We inquired \smabb{C} about the toughest phase of the security lifecycle (i.e., \textit{prevention}, \textit{detection}, \textit{reaction}), and report their answers in Fig.~\ref{fig:phases}: we can see that 42\% of \smabb{C} consider \textit{prevention} to be the toughest stage of the security lifecycle---while 28\% consider \textit{detection} to be harder.\footnote{Interestingly, two companies provided an independent answer to such a question: one stated that ``all phases are equally challenging,'' and another stated that ``the real challenge is determining which assets (and threats) should be prioritized.''} We asked the same question also to \smabb{A}: accordingly, 100\% of \smabb{A} believe that it is the \textit{detection} phase that is the hardest for \smabb{C}.

\vspace{-1em}


\subsection{Education and Information Sharing (\smabb{C} \& \smabb{A})}
\label{ssec:education}
One of the greatest barriers towards the creation of secure systems is convincing executives of the importance of cybersecurity~\cite{apruzzese2022role}. Furthermore, it is  well-known that employees are the weakest link in the security chain. Hence, we asked \smabb{C} to provide a rough estimate about the \textit{awareness} of cybersecurity across their organizations, and report the results in Table~\ref{tab:awareness}. According to this table, the impression is that the overall quality of awareness/education is satisfactory for all \smabb{C}---although there is still room for improvement.

\begin{table}[!htbp]
    \centering
    \caption{Cybersecurity \textit{awareness} and \textit{education} across diverse organisational levels---as reported by \scbb{C}. }
    \label{tab:awareness}
    \resizebox{0.95\columnwidth}{!}{
        \begin{tabular}{|l|c|}
            \hline
        	\multicolumn{2}{|c|}{\textbf{Mid-/Top-level management}}\\
        	\hline
        	\textit{Option} & \textit{Freq.} \\
        	\hline
        	They are fully aware of the risks and prioritise cyber-security & 64.29\% \\
        	They are fully aware of the risks, but cyber-security is not a priority & 21.43\% \\
        	They are not aware of the risks, but are educated on the topic & 7.14\% \\
        	No answer & 7.14\% \\
        	\hline \hline
        	
        	\hline
        	\multicolumn{2}{|c|}{\textbf{Employees}}\\
        	\hline
        	\textit{Option} & \textit{Freq.} \\
        	\hline
        	They are aware fully of the risks and education is evaluated regularly & 50.00\% \\
        	They are not fully aware of the risks, but are educated on the topic & 42.86\% \\
        	They are not aware of the risks, and unlikely to improve in the short-term & 0.00\% \\
        	No answer & 7.14\% \\
        	
        	\hline
        	
        \end{tabular}
    }
\end{table}

\begin{figure}[!htbp]
    \centering
    \includegraphics[width=0.65\columnwidth]{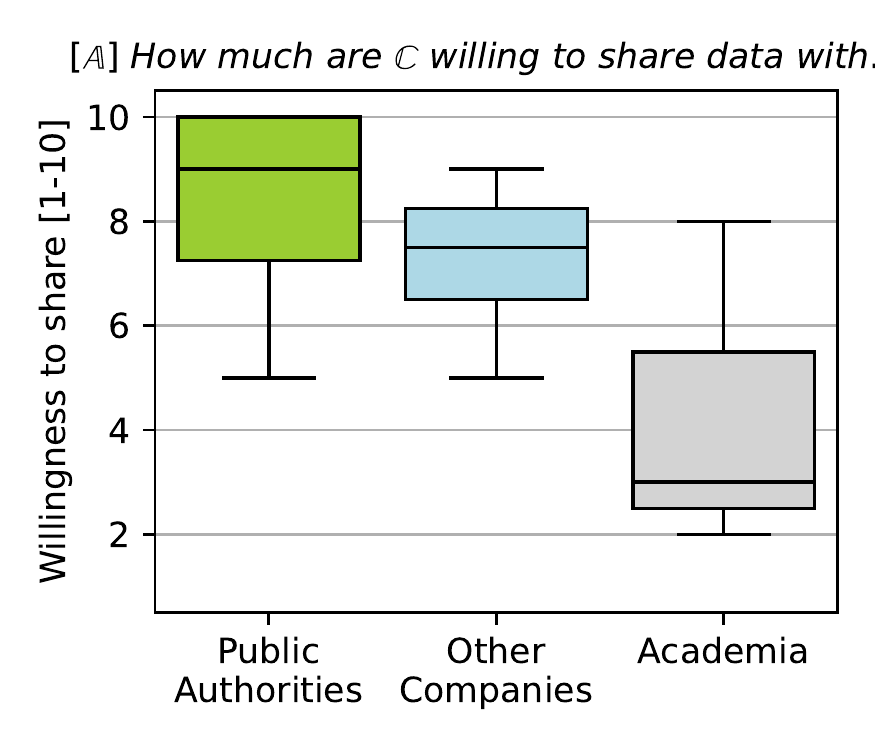}
    \vspace{-0.5em}
    \caption{Perceived (from \scbb{A}) willingness of \scbb{C} to share information with other entities---on a [1--10] scale.}
    \label{fig:sharing}
    \vspace{-1em}
\end{figure}

We also investigated the importance of \textit{information sharing} from the perspective of both \smabb{C} and \smabb{A}. Indeed, a \smabb{C} can improve their cybersecurity by studying attacks targeting other \smabb{C}: 78\% of \smabb{C} already do this (the remaining 22\% plans to). With respect to \smabb{A}, we inquired about their perceived willingness (on a [1--10] scale) of \smabb{C} to disclose information to other entities, namely: academia, different \smabb{C}, and \smabb{A}. The answers, in Fig.~\ref{fig:sharing}, confirm our own impressions (§\ref{ssec:challenges}): \smabb{C} are reluctant to cooperate with scientific researchers.

\vspace{-0.5em}

\subsection{Risk Assessment}
\label{ssec:risk}

The SG can be targeted by a plethora of cyberthreats (§\ref{sec:related}). Our survey covers such topic from the twofold perspective of \textit{cyber-} and \textit{physical}-risk. When inquired about how such risks are assessed, all \smabb{C} responded that they rely on \textit{qualitative} metrics---whereas quantitative metrics are often neglected.

\begin{figure}[!htbp]
    \centering
    \includegraphics[width=0.75\columnwidth]{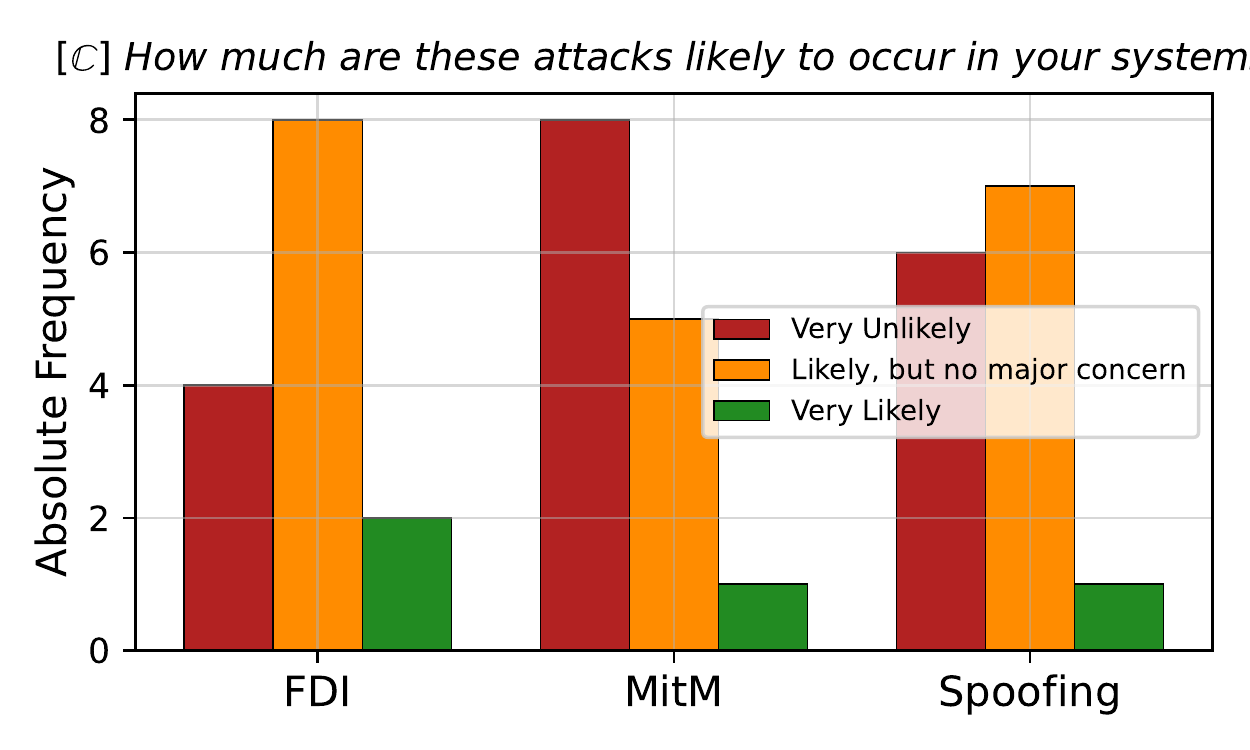}
    \caption{Threat of FDI, MitM, and Spoofing attacks (for \scbb{C}).}
    \label{fig:attacks_companies}
    \vspace{-0.5em}
\end{figure}

\textbf{Cyber-risk.}
All \smabb{C} consider their systems to be at risk from \textit{APT}. Moreover, only 14\% of \smabb{C} consider illegitimate \textit{access to consumer data} to be `not threatening', and none of \smabb{C} consider \textit{DoS} to be problematic. We report in Fig.~\ref{fig:attacks_companies} the perceived risk (according to \smabb{C}) of well-known cyberattacks, namely: FDI, MitM, spoofing. 

\begin{figure}[!htbp]
    \centering
    \includegraphics[width=0.7\columnwidth]{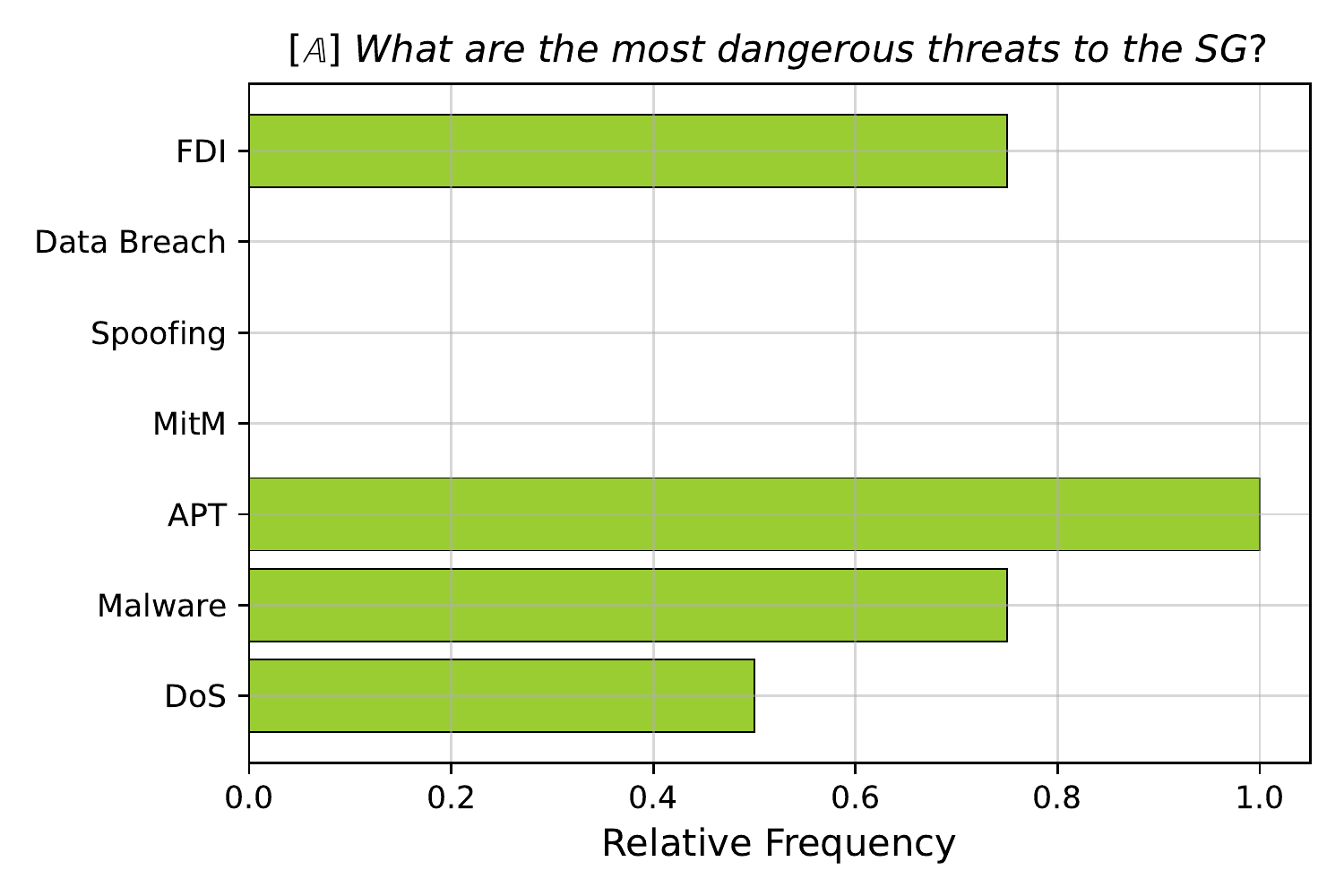}
    \caption{Most dangerous threats to the SG (according to \scbb{A}).}
    \label{fig:attacks_authorities}
    \vspace{-1em}
\end{figure}

We asked similar questions (i.e., ``how feasible are these cyberthreats to the SG?'') to \smabb{A}, whose responses are shown in Fig.~\ref{fig:attacks_authorities}. We also inquired \smabb{A} to estimate the cybersecurity capabilities of \smabb{C}. Accordingly, none of \smabb{A} consider \smabb{C} to have `excellent' cybersecurity measures in terms of \textit{prevention} or \textit{detection}; however, 25\% of \smabb{A} considers \smabb{C} to have excellent \textit{reaction} capabilities.

\begin{figure}[!htbp]
    \centering
    \includegraphics[width=1\columnwidth]{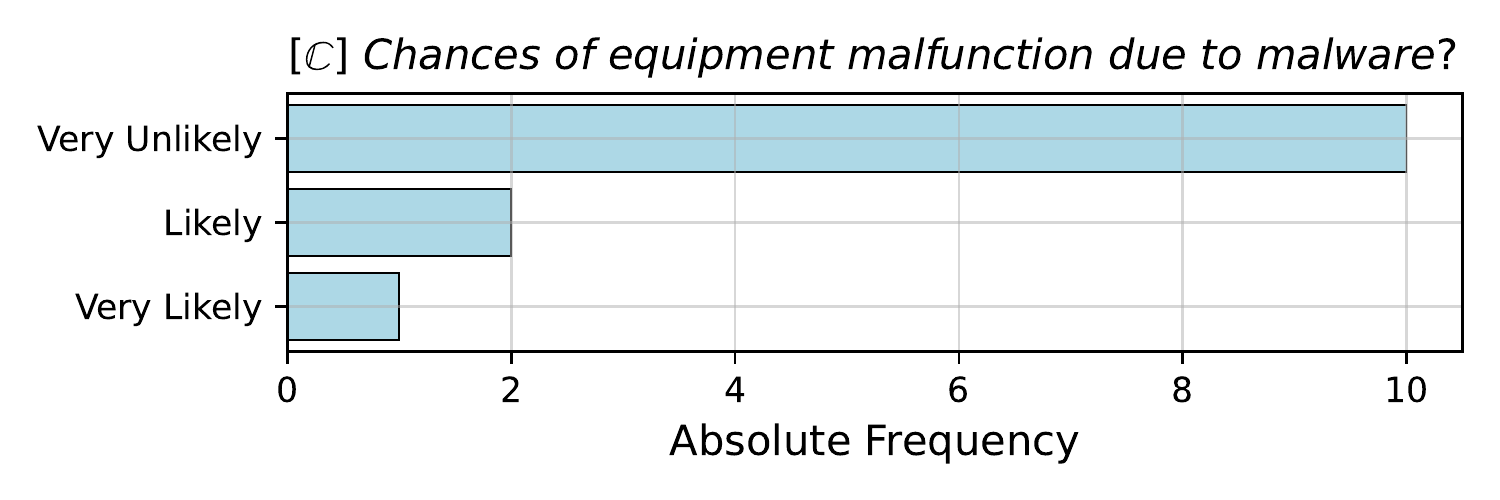}
    \caption{Opinion on \scbb{C} on the likelihood of malware inducing equipment malfunction.}
    \label{fig:malfunction}
    \vspace{-1em}
\end{figure}

\textbf{Physical-risk.}
None of \smabb{C} consider `physical' security to be \textit{more important} than `cyber' security. As a matter of fact, when asked about the likelihood of malware inducing equipment malfunction \textit{harmful to humans} (shown in Fig.~\ref{fig:malfunction}), 70\% of \smabb{C} stated that it is ``very unlikely.'' Moreover, inspired by a 2021 report from Gartner~\cite{gartner2021kill} announcing the term ``killware'', we asked both \smabb{C} and \smabb{A} whether it was possible that ``malware can lead to human death'' in SG. Results differed between \smabb{C} and \smabb{A}: 14\% of \smabb{C} believe that such a risk to be completely unrealistic, and 71\% think that it is unlikely. In contrast, 50\% of \smabb{A} agree that ``killware'' is a very likely threat in SG, and none believe that it is unrealistic. 

\subsection{Technological Paradigms}
\label{ssec:paradigms}
Due to its increasing reliance on IT, the SG is an enticing setting for modern technological paradigms~\cite{kumar2020distributed}, such as Blockchain, Artificial Intelligence~(AI), Internet of Things~(IoT), or Cloud Computing. 

\begin{figure}[!htbp]
    \centering
    \includegraphics[width=0.85\columnwidth]{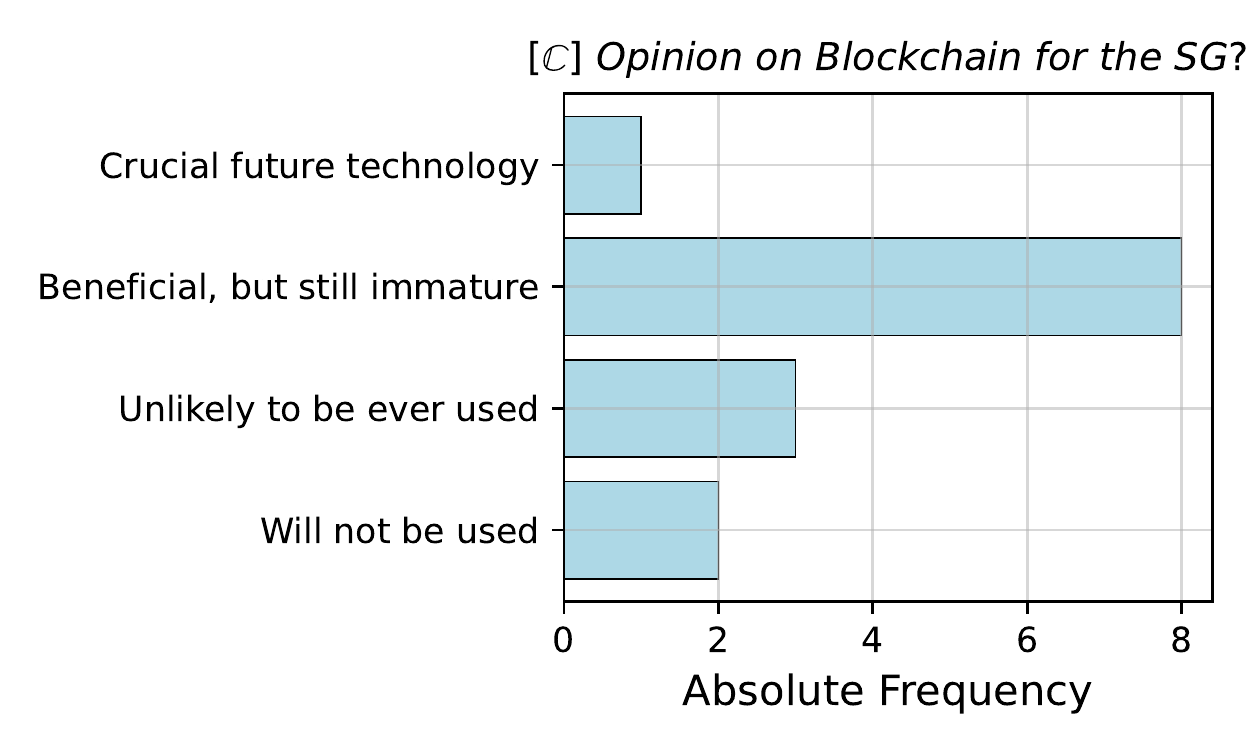}
    \caption{Opinion of \scbb{C} on the utility of \textit{blockchain} for SG.}
    \label{fig:blockchain}
\end{figure}

We first asked \smabb{C} about their perspective on similar technologies, starting from their opinion on \textit{Blockchain} for the SG. The answers, in Fig.~\ref{fig:blockchain}, suggest that \smabb{C} does not hold \textit{Blockchain} in high-regards.

\begin{figure}[!htbp]
    \centering
    \includegraphics[width=1\columnwidth]{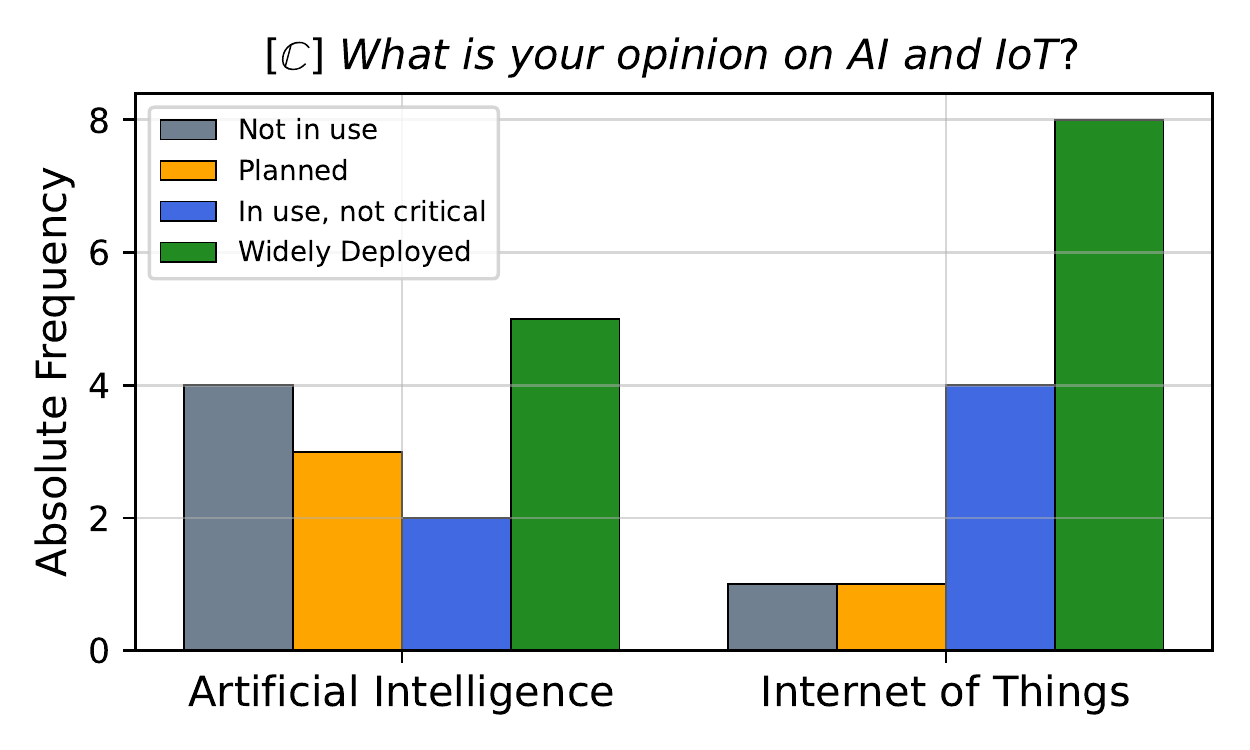}
    \caption{Answers of \scbb{C} on their adoption of \textit{AI} and of \textit{IoT}.}
    \label{fig:aiiot_companies}
\end{figure}

With regards to AI and IoT, we report the answers of \smabb{C} in Fig.~\ref{fig:aiiot_companies}, showing a wide adoption of \textit{IoT} (only 14\% do not use it), a much smaller one for \textit{AI} (50\% do not use it). Lastly, we report that only 7\% of \smabb{C} does not use \textit{Cloud} solutions.


\begin{figure}[!htbp]
    \centering
    \includegraphics[width=0.8\columnwidth]{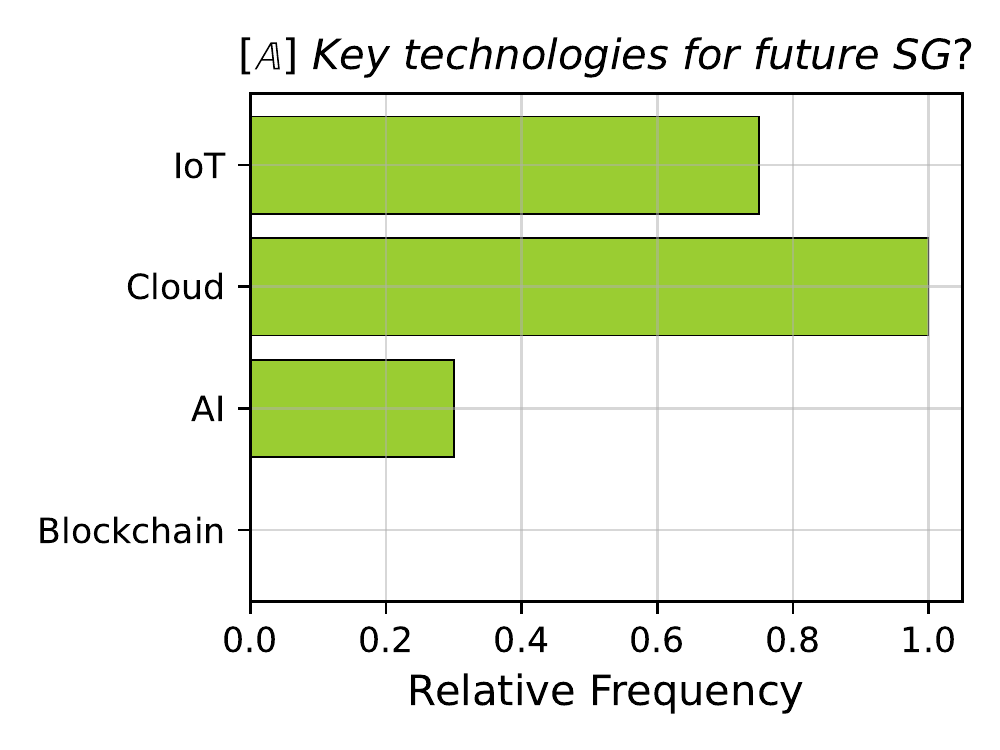}
    \caption{The perspective of \scbb{A} on the technologies that will have a biggest impact in the future for cybersecurity in SG.}
    \label{fig:future_authorities}
    
\end{figure}

We then asked \smabb{A} about their opinion on the relevance of all such paradigms for the SG, and report their answers in Fig.~\ref{fig:future_authorities}.


Finally, we report that 86\% of \smabb{C} use \textit{smart meters}~\cite{el2018smartgrid}, while the remaining 14\% plans to deploy them in the short-term.

\section{Analysis (contrast)}
\label{sec:analysis}
After a transparent presentation of our findings, we now provide a deeper and \textit{objective} analysis addressed at elucidating some blind spots---which may inspire constructive developments of cybersecurity in the SG. Specifically, we focus on the \textit{points of contrast} that emerge from our survey. 
We first consider the viewpoint of practitioners w.r.t. those of researchers~(§\ref{ssec:pravres}), and then focus on the contrasting opinions of private companies and public authorities~(§\ref{ssec:pubvpriv}). Finally, we make some considerations on our study~(§\ref{ssec:considerations}).

\subsection{Practitioners vs Researchers}
\label{ssec:pravres}
Our survey highlights a stark disconnection between the perspective of practitioners (i.e.,~\smabb{C} and \smabb{A}) and researchers---at least according to what is reported in scientific literature related to the SG. 
In particular, we identify five major points of divergence.

\begin{enumerate}[label={\small {{\roman*}})}]
    \item \textbf{Blockchain.} 
    Both \smabb{C} and \smabb{A} agree that blockchain technologies are not very significant in the SG context (see Fig.~\ref{fig:blockchain} and Fig.~\ref{fig:future_authorities}). However, many recent research papers endorse the application of blockchain in the SG (e.g.,~\cite{mollah2020blockchain, kumar2020distributed, PC_zhao2022lightweight, PC_chatzigiannis2021black}). 
    
    \item \textbf{Artificial Intelligence.} 
    Only half of \smabb{C} uses AI solutions (Fig.~\ref{fig:aiiot_companies}), whereas \smabb{A} does not believe that AI will have much impact in the future of the SG (Fig.~\ref{fig:future_authorities}). In contrast, recent papers (e.g.,~\cite{kumar2020distributed, kshetri2021economics, apruzzese2022role}) claim that AI will (or already does) play a substantial role in critical infrastructures---especially from a cybersecurity standpoint (e.g., detecting cyberthreats~\cite{PC_kwon2019rnn}). 
    
    \item \textbf{Dangerous Cyberthreats.}
    Most \smabb{C} do not believe that \textit{MitM} or \textit{spoofing} attacks are dangerous to their systems (Fig.~\ref{fig:attacks_companies}), which is an opinion also shared by \smabb{A} (Fig.~\ref{fig:attacks_authorities}). However, abundant (and recent) research papers claim that MitM and spoofing are extremely dangerous to the SG (e.g.~\cite{chauhan2021synchrophasor, PC_wlazlo2021man}).
    
    \item \textbf{Risk Assessment Methods.}
    All \smabb{C} adopt \textit{qualitative} approaches for their risk assessment (§\ref{ssec:main}). In contrast, such methods are only marginally considered in scientific literature, which mostly focuses on \textit{quantitative} methods~\cite{PC_young2016framework}.
    
    \item \textbf{Reaction Phase.}
    Most \smabb{C} primarily focus on \textit{reaction} to cyberthreats, due to the acknowledged difficulty in preventing/detecting them. However, in research it is the complete opposite: very few papers focus on \textit{reaction}~\cite{PC_salazar2019enhancing}. 
\end{enumerate}
(Citing \textit{all} research papers that support these points of divergence is outside our scope---and is also unfeasible in this single publication.)

Given its importance, we also emphasize the viewpoint of practitioners and research with regards to \textbf{malware leading to physical harm.} Specifically, both \smabb{A} and \smabb{C} tend to agree that ``killware''~\cite{gartner2021kill} is a possibility---albeit not very likely (§\ref{ssec:risk}). In contrast, some very recent works already used the term ``killware'' (e.g.,~\cite{tuttle20222022, bowers2022securing}), whereas many research papers (e.g.,~\cite{dissanayake2021grounded}) are citing~\cite{eddy2020cyber} to claim that exploiting cyber-vulnerabilities can lead to ``human death''. (According to Google Scholar,~\cite{eddy2020cyber} has 30 citations as of Sept. 2022.)

\begin{cooltextbox}
\textbf{\textsc{Observation:}}
there is a relevant mismatch between the opinions of \textit{practitioners} and the focus of \textit{research} papers in the context of SG cybersecurity.
\end{cooltextbox}

\subsection{Private Companies vs Public Authorities}
\label{ssec:pubvpriv}
We find instructive to compare the viewpoints of the \textit{private} (\smabb{C}) and \textit{public} (\smabb{A}) sector. 
Despite \smabb{C} and \smabb{A} agree on some aspects (e.g., which technological paradigms will play a pivotal role in the future SG,~§\ref{ssec:paradigms}), some differences emerge when comparing some responses of \smabb{C} against those of \smabb{A}. Let us illustrate five significant ones.

\begin{enumerate}[label={\small {{\roman*}})}]
    \item \textbf{Detection phase.}
    Although all \smabb{A} consider \textit{detection} to be the toughest phase of cybersecurity, such opinion is shared only by 28\% of~\smabb{C} (§\ref{ssec:main}). 

    \item \textbf{Capabilities.}
    Most \smabb{C} believe to be \textit{well-equipped} in terms of cybersecurity (only APT, apparently, endanger \smabb{C}); \smabb{A}, however, have a significantly different opinion (§\ref{ssec:risk}). 
    
    \item \textbf{Data Confidentiality.}
    Most \smabb{C} consider \textit{data confidentiality} (§\ref{ssec:main}) to be a problem, while \smabb{A} believe that data breaches are not a severe threat to the SG (Fig.~\ref{fig:attacks_authorities}). 
    
    \item \textbf{Data Replication.}
    All \smabb{C} adopt \textit{data replication} strategies---sometimes by leveraging cloud solutions (§\ref{ssec:main}). In contrast, some \smabb{A} believe that such strategies should be discouraged. 
    
    \item \textbf{FDI and DDoS.}
    While \smabb{C} do not consider FDI or DDoS attacks to be problematic, \smabb{A} frequently named them among the most threatening scenarios (§\ref{ssec:risk}).  
\end{enumerate}

\begin{cooltextbox}
\textbf{\textsc{Observation:}}
The \textit{public} and \textit{private} sectors have some diverging opinions on the current state of SG cybersecurity.
\end{cooltextbox}

\subsection{Considerations}
\label{ssec:considerations}

Let us reflect on three aspects of our study, and then provide a disclaimer to prevent generating harmful misunderstandings.

\textbf{(1) Focus.} 
Our goal is providing a holistic overview of cybersecurity in the entire SG. Therefore, deriving conclusions that apply for individual domains of the SG (e.g., electricity market~\cite{tesfamicael2020design}) or for specific components deployed in such domains (e.g., CPS~\cite{PC_aoudi2018truth} or ICS~\cite{line2013case}) is outside our scope. Nonetheless, the entities interviewed in our study pertain to \textit{all segments of the SG}, all of which employ state-of-the-art components and practices in their infrastructures.

\textbf{(2) Area.}
Our study focuses on entities operating \textit{in Europe}. Hence, some of our findings may not apply to SG located in different areas of the world (e.g., the USA). However, \textit{research is universal}: a novel solution proposed in a scientific paper can very well be deployed in a country that is different from that of the paper's authors, or of the paper's publication venue. This is the reason why our analysis (§\ref{ssec:pravres}) is not confined to `European' papers.

\textbf{(3) Population.}
Our study covers 18 entities, spanning across both private companies and public authorities. Such a number may appear small compared to, e.g., the interviews carried out by some papers (e.g., the one by Grosse et al.~\cite{grosse2022so}). However, our population is comparable to the one in~\cite{line2013case} in terms of raw-numbers. Moreover, our interviews reflect the viewpoint of (i) public authorities that represent entire countries---having \textit{millions} of citizens; and (ii) private companies, some of which count \textit{thousands of employees} and large IT departments. Furthermore, a given country in Europe may have only one or two companies that focus on power generation---and such companies can also operate in other countries within Europe. Therefore, our interviews allows to meet our research objective: investigating the current state of cybersecurity in the European SG.

\textbf{Disclaimer.}
Insofar, the purpose of our study was providing an \textit{objective} assessment of the SG's practitioners' viewpoint on some cybersecurity aspects.
We do not make any general claim, nor attempt to make any hypotheses that will be subject to `statistical testings'. It is unfair (and dangerous) to derive, e.g., that ``MitM attacks are not important for practitioners'' (cf. Fig.~\ref{fig:attacks_companies}): despite being a minority, \textit{some} companies feel endangered by MitM, and hence it is important that some researchers continue to study such threats.
\section{Interpretation (original)}
\label{sec:interpretation}
As a last contribution of this paper, we provide an original interpretation of our study. Our intention is twofold: (i)~\textit{explaining} the reasons behind some of our findings; and (ii)~deriving \textit{actionable recommendations} for future works.

To this purpose, we rely on the abundant notes we took during our interviews, and combine them with our own expertise. Hence, in the remainder of this section, we: provide our explanations for each identified point of \textit{contrast} (presented in §\ref{sec:analysis}) in Table~\ref{tab:recommendations}; present how \textit{regulations} affect the cybersecurity in the SG~(§\ref{ssec:regulations}); and highlight the major \textit{takeaways}~(§\ref{ssec:takeaways}) that can lead to significant improvements of cybersecurity in the SG.

\begin{table*}[!htbp]
    \centering
    \caption{Points of disagreement and our comment.} 
    \label{tab:recommendations}
        \begin{tabularx}{2\columnwidth}{c|l|X}
             \toprule
             \textbf{Sphere} & \textbf{Disagreement} & \textbf{Our Explanation/Recommendation} \\
             \midrule
             \multirow{21}{*}{\begin{tabular}{c}
                 Research \\ 
                 vs \\
                 Practice \\
                 (§\ref{ssec:pravres})
             \end{tabular} } & Blockchain & {\small Our explanation is that either researchers \textit{overestimate} the real-world impact of blockchain; or the benefits of blockchain are not communicated in a \textit{convincing way} (or both). Nevertheless, a huge barrier is represented by the difficulty~\cite{kayikci2022critical} in implementing real blockchain systems \textit{in practice}.} \\ \cmidrule{2-3}
             
             & Artificial Intelligence & {\small We believe the poor consideration of AI by practitioners stems from two reasons: the impending \textit{regulations}~\cite{Europe:AI}, which make it difficult for practitioners to setup and use AI in the SG; and the lack of \textit{explainability}~\cite{apruzzese2022role} of most AI methods which prevent determining the root cause of `failures' in AI systems (recall that cybersecurity is mostly outsourced in the SG §\ref{sec:findings}).} \\ \cmidrule{2-3}
             
             & MitM and Spoofing & {\small We endorse research to keep investigating also MitM and spoofing attacks. However, such threats should be exemplified in the \textit{broad context of APTs} targeting the SG, since they are of major interest for practitioners (e.g., MitM can very well be part of APT~\cite{park2021advanced}) .} \\ \cmidrule{2-3}
             
             & Qualitative Assessments & {\small We conjecture that the lack of qualitative assessments \textit{in research} is due to the difficulty in devising a `scientifically sound' and `generally applicable' qualitative assessment method---whose application \textit{in practice} is case-specific, and present high-degrees of subjectivity~\cite{touhiduzzaman2019review}.} \\ \cmidrule{2-3}
             
             & Reaction phase & {\small The \textit{reaction} phase is difficult to model \textit{in research}, because \textit{in practice} such phase is tailored for the specific asset involved in the cyberattack. Nonetheless, we endorse future research to put more emphasis on this phase because improving its efficiency allows practitioners to invest more resources in other---tougher---phases (e.g., detection).}\\ \cmidrule{2-3}
             
             & ``Killware'' & {\small Given the sensitivity of this subject, we endorse future research to \textit{treat it with care}. In particular, we recommend to avoid using such term to ``sell'' a given paper: it is well-known that cyberattacks are dangerous (especially in the SG), and mentioning a problem that (for the time being) is just a remote possibility can cause more harm than good.} \\

             \midrule \midrule 
             
             \multirow{17}{*}{\begin{tabular}{c}
                 Public \\
                 vs \\
                 Private \\
                 (§\ref{ssec:pubvpriv})
             \end{tabular} } & Prevention phase & {\small We conjecture that \smabb{A} consider \textit{detection} to be more challenging for \smabb{C} due to the well-known difficulty in adopting significant \textit{prevention} measures. Hence, it is likely that \smabb{A} either: did not consider `prevention' to be a meaningful answer; or a phase actively pursued by \smabb{C}.} \\ \cmidrule{2-3}
             
             & Capabilities & {\small Two possibilities can explain why the viewpoint of \smabb{A} does not align with that of \smabb{C} in terms of overall cybersecurity capabilities: either \smabb{C} \textit{overestimate} their cybersecurity (either their own, or the outsourced one); or \smabb{A} \textit{underestimates} the capabilities of \smabb{C} (or both of these). Nevertheless, a better cooperation between \smabb{C} and \smabb{A} can allow to reach a consensus.} \\ \cmidrule{2-3}
             
             & Data Confidentiality & {\small We believe that \smabb{A} does not consider data confidentiality to be a problem because \textit{privacy violations do not lead to outages}. In contrast, such violations are a big threat to \smabb{C}, because to \textit{comply with regulations} they must protect their customers' data---which is challenging.} \\ \cmidrule{2-3}
             
             & Data Replication & {\small The reason why \smabb{A} discourage such strategies (commonly used by \smabb{C}) is that data replication inevitably extends the surface that can be exploited by attackers to access private data. Although the concerns of \smabb{A} are well-founded, we believe that---in the SG---privacy is not as critical as reliable functionality of the entire network. Nonetheless, future works can improve such privacy.}\\ \cmidrule{2-3}
             
             & FDI and DDoS & {\small A possible explanation as to why \smabb{A} considers FDI or DDoS to be threatening is that such attacks are very popular in papers and reports (e.g.,~\cite{liang2016review, Gartner:Cybersecurity}), thereby increasing their perceived threat (according to \smabb{A}). However, \smabb{C} are well-aware of whether their systems are at risk of FDI or DDoS. A better information exchange between \smabb{A} and \smabb{C} could lead to a more even distribution of opinions.}\\
            \bottomrule             
        \end{tabularx}
\end{table*}

\subsection{The role of Regulations in the SG}
\label{ssec:regulations}
A recurrent theme during our interviews with \smabb{C} involved the role of regulations the the cybersecurity of the SG.

\textbf{A hidden issue.} 
Most \smabb{C} revealed that one of the greatest challenges for companies (not just their own) is represented by the constantly mutating and often unclear regulations that \smabb{C} must abide to. Such a situation is indeed problematic: ultimately, \smabb{C} are \textit{businesses}, with a finite amount of resources. Hence, if the regulations are difficult to interpret, \smabb{C} either (a) delay their application---perhaps until after other \smabb{C} found a way to implement them; or (b) must invest resources in legal consultants---resources that could be spent in improving their own systems. It then follows that such regulations---despite representing a protection mechanism \textit{in principle}---can be detrimental \textit{in practice}. For example, consider the divergent opinion of \smabb{A} and \smabb{C} with regards to data confidentiality (§\ref{ssec:pubvpriv}). \smabb{A} state that it is not a big concern---which makes sense, as there are far more disruptive hazards that can stem from failures in the SG. However, \smabb{C} report that ensuring that all data are kept secure is very tough---because \smabb{C} \textit{must} ensure such confidentiality to comply with regulations. Put differently: \smabb{A}---who must oversee compliance with regulations by \smabb{C}---lack the operational perspective of \smabb{C} and hence overlook most of the difficulties originating from such regulations.

\begin{figure}[!htbp]
    \centering
    \includegraphics[width=1\columnwidth]{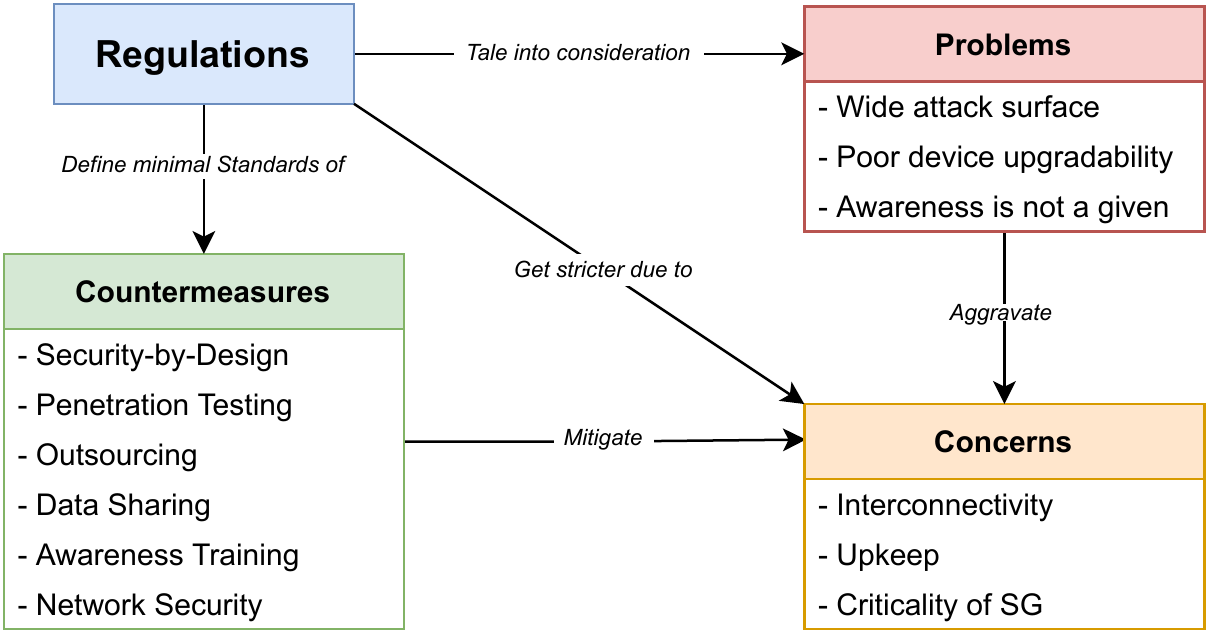}
    \caption{Our original model displaying the relationships between \textit{regulations} the cybersecurity of the SG.}
    \label{fig:interpretation}
\end{figure}

\textbf{Our Model.}
To exemplify how regulations affect the current (and future) cybersecurity in the SG, we review all our notes, abstract them, and then derive an original model shown in Fig.~\ref{fig:interpretation}.
Such model elucidates the relationships between the regulations (i.e., the legal foundations that bind \smabb{C}) and: the \textit{concerns} of practitioners (both \smabb{C} and \smabb{A}) regarding current developments; the major cybersecurity \textit{problems} in SG; and the potential \textit{countermeasures} to face such problems. 

Let us describe Fig.~\ref{fig:interpretation}. The concerns of practitioners are aggravated by the problems posed by current developments in IT related to the SG. At the same time, such concerns are mitigated by the fact that effective countermeasures are available. 
\textbf{Legislative bodies} then consider (a)~the current problems defined by the constantly evolving threat landscape, and (b)~how such problems can be addressed via existing countermeasures; and enact `stricter' regulations that simultaneously (i)~aim to better protect the SG, and (ii)~fuel the concerns of private companies operating the SG.


\subsection{Takeaways}
\label{ssec:takeaways}
In this paper, we mentioned four distinct `spheres': (i)~private companies, (ii)~public authorities, (iii)~research and academia, and (iv)~legislative bodies. By aggregating all our findings and interpretations, we now derive four takeaways---one for each of these spheres.

\begin{itemize}
    \item \textbf{Private companies} should be \textit{more open} in cooperating with research institutions. The latter can greatly benefit \smabb{C} (by, e.g., devising new standards \cite{krotofil2019securing}, or developing novel devices~\cite{PC_werth2020prototyping})---but significant developments cannot be expected unless \smabb{C} become more willing to share some of their concerns with academia.\footnote{We observe that a similar message was sent by Awad et al.\cite{PC_awad2018tools} in 2018: unfortunately, as demonstrated by our own difficulties (§\ref{ssec:challenges}), the problem is still open.}

    \item \textbf{Research and academia} should account for the \textit{viewpoint of practitioners} (i.e., \smabb{A} but especially \smabb{C}) by prioritizing the most likely (and dangerous) threats to the SG, avoid over-exaggerations, and better communicate (by, e.g., using economical terms) the benefit of novel technologies. Moreover, future research should not neglect the reaction phase, as it can greatly benefit operational cybersecurity.
    
    \item \textbf{Legislative bodies} should enact \textit{actionable and stable regulations}. \smabb{C} are often overwhelmed, leading to `closeness' w.r.t. information sharing (due to, e.g., privacy), as well as to `waste` of resources to interpret the ever-mutating rules.
    
    \item \textbf{Public authorities} should better engage with private companies. Indeed, \smabb{C} `must' respond to \smabb{A}, and hence \smabb{A} should use this opportunity to better understand the concerns of \smabb{C} and serve as the \textit{glue} between all the other spheres, as well as with the citizens of their respective countries.
\end{itemize}

\begin{cooltextbox}
\textbf{\textsc{Recommendation:}}
we endorse all such spheres to better communicate and interact with each other. Ultimately, they have a common goal: improving the security of the SG.
\end{cooltextbox}
\section{Conclusions}
\label{sec:conclusions}
Our paper aims to elucidate the `internal' perspective of the Smart Grid (SG) with respect to cybersecurity. To this purpose, we carry out and present the results of a large survey conducted with practitioners, spanning over both \textit{private companies} and \textit{public authorities} from diverse countries in Europe. After objectively presenting our major results and highlighting some contrasting points of disagreement, we provide our interpretation of our findings---casting light on some concealed problems that impair a smooth development of the SG.
Our paper will hopefully inspire future endeavors that -- by improving the security of the Smart Grid -- will pave the way to a reliable growth of our society.

\textbf{Acknowledgement.} We thank Katherine Grosse for some early feedback on this paper; and the Hilti Corporation for funding.



\end{document}